\documentclass[aps,pre,superscriptaddress,twocolumn,showpacs,floatfix]{revtex4-1}
\usepackage{graphicx}

\begin{document}


\title{Intrinsic core level photoemission of suspended  graphene}


\author{Toma Susi}
\email[]{Author to whom correspondence should be addressed: toma.susi@univie.ac.at}
\affiliation{University of Vienna, Faculty of Physics, Boltzmanngasse 5, 1090 Vienna, Austria}

\author{Mattia Scardamaglia}
\affiliation{Chemistry of Interaction Plasma-Surface (ChIPS), University of Mons, Mons, Belgium}

\author{Kimmo Mustonen}
\author{Andreas Mittelberger}
\affiliation{University of Vienna, Faculty of Physics, Boltzmanngasse 5, 1090 Vienna, Austria}

\author{Mohamed Al-Hada}
\affiliation{Elettra - Sincrotrone Trieste ScPA, Area Science Park, 34149 Basovizza-Trieste, Italy}
\affiliation{Department of Physics, College of Education and Linguistics, University of Amran, Yemen}
\author{Matteo Amati}
\author{Hikmet Sezen}
\author{Patrick Zeller}
\affiliation{Elettra - Sincrotrone Trieste ScPA, Area Science Park, 34149 Basovizza-Trieste, Italy}

\author{Ask H. Larsen}
\affiliation{Nano-bio Spectroscopy Group and ETSF Scientific Development Centre, Departamento de F\'isica\\ de Materiales
Universidad del Pa\'is Vasco UPV/EHU, Av. Tolosa 72, E-20018 San Sebasti\'an, Spain}

\author{Clemens Mangler}
\author{Jannik C. Meyer}
\affiliation{University of Vienna, Faculty of Physics, Boltzmanngasse 5, 1090 Vienna, Austria}

\author{Luca Gregoratti}
\affiliation{Elettra - Sincrotrone Trieste ScPA, Area Science Park, 34149 Basovizza-Trieste, Italy}

\author{Carla Bittencourt}
\affiliation{Chemistry of Interaction Plasma-Surface (ChIPS), University of Mons, Mons, Belgium}

\author{Jani Kotakoski}
\affiliation{University of Vienna, Faculty of Physics, Boltzmanngasse 5, 1090 Vienna, Austria}

\date{\today}

\begin{abstract}
X-ray photoelectron spectroscopy of graphene is important both for its characterization and as a model for other carbon materials. Despite great recent interest, the intrinsic photoemission of its single layer has not been unambiguously measured, nor is the layer-dependence in free-standing multilayers accurately determined. We combine scanning transmission electron microscopy and Raman spectroscopy with synchrotron-based scanning photoelectron microscopy to characterize the same areas of suspended graphene samples down to the atomic level. This allows us to assign spectral signals to regions of precisely known layer number and purity. The core level binding energy of the monolayer is measured at 284.70~eV, thus 0.28~eV higher than that of graphite, with intermediate values found for few layers. This trend is reproduced by density functional theory with or without explicit van der Waals interactions, indicating that intralayer charge rearrangement dominates, but in our model of static screening the magnitudes of the shifts are underestimated by half.

\vspace{30 pt}
\end{abstract}


\maketitle

In addition to providing an elemental fingerprint, the kinetic energies of electrons ejected from the core levels of atoms carry information on their local bonding and dielectric environment due to the screening of the core hole during the photoemission process. X-ray photoelectron spectroscopy (XPS) can thus be used to study the surface composition of materials, and is a powerful probe of the chemical and electronic structure of low-dimensional carbon nanomaterials~\cite{Susi15BJN}, including nanotubes~\cite{Ayala09PRB} and graphene~\cite{Hibino09PRB,Lizzit10NP}. Besides its superb properties and potential technological relevance, graphene is an useful model system for the physics of photoemission.

Since core electrons are localized and do not participate in chemical bonding, narrow linewidths of core level signals can be expected. However, in graphene the C~1\textit{s} levels do show some dispersion~\cite{Lizzit10NP}, and more importantly, their photoemission signal has significant asymmetry towards higher binding energies due to excitation of low-energy electron-hole pairs screening the core hole in metallic systems~\cite{Doniach70JoPC}. The C 1\textit{s} core level binding energy (BE) of graphite is known to be 284.42 eV~\cite{Prince00PRB, Lizzit07PRB, Balasubramanian01PRB} (photoelectron signal with a 160--180 meV lifetime broadening and an asymmetry of 0.05--0.065~\cite{Prince00PRB, Lizzit07PRB}). A second occasionally measured component shifted to higher BEs by 120--194 meV has been attributed to the surface layer~\cite{Balasubramanian01PRB, Hunt08PRB}, but this has been disputed~\cite{Lizzit07PRB, Lizzit10NP}.

On metallic substrates where graphene is typically grown, large Dirac point variations~\cite{Niesner14JPCM} lead to monolayer BEs ranging from 283.97 eV on Pt(111)~\cite{Preobrajenski08PRB} to 284.7 eV on Ni(111)~\cite{Grueneis09NJP}, and for epitaxial graphene on SiC, the monolayer value was found to be $\sim$0.4 eV upshifted from that of a four-layer area~\cite{Hibino09PRB}. It is thus clear that the environment significantly affects the BE. A measurement of its intrinsic value for suspended monolayer graphene, only possible with spatially resolved XPS, has to date not been performed, hampering metrology and efforts to further increase the precision of modeling~\cite{Susi15PRB}.

We combine synchrotron-based scanning photoelectron microscopy (SPEM) with atomic resolution scanning transmission electron microscopy (STEM) and Raman spectroscopy to comprehensively characterize suspended graphene areas, and to measure their high-resolution XPS spectra. We find the intrinsic BE of the monolayer to be 284.70$\pm$0.05 eV, with bilayer and four-layer found respectively at 284.54 and 284.47~eV. Regardless of whether van der Waals interactions are explicitly included, density functional theory correctly predicts this trend, but underestimates the magnitude of the shifts.

Our first graphene sample was synthesized by chemical vapor deposition (Quantifoil\textregistered~R~2/4, Graphenea), and the second one by mechanical exfoliation,  both transferred onto gold support grids with perforated carbon membranes. The first sample contains a good coverage of mostly monolayer (1L) graphene, with occasional grain boundaries and small multilayer grains, while the second one has regions of variable layer thickness. To clean the samples from residual contamination, we used vacuum laser annealing, capable of producing samples with atomically clean areas spanning several hundreds of nm$^2$~\cite{Tripathi17PSSRRL}. The samples were further annealed in vacuum at 500~$^{\circ}$C prior to the spectromicroscopy experiment to reduce any contamination absorbed during the ambient transfer.

The BEs were measured using X-ray SPEM at the Escamicroscopy beamline of the ELETTRA synchrotron~\cite{Abyaneh11eSSN}. The suspended graphene areas were first located by imaging, and spectra then collected from 130~nm spots given by the demagnifying action of a Fresnel zone plate on the X-ray beam produced by the synchrotron storage ring. The photon energy was 401.03~eV (Au~4\textit{f} reference), with an energy resolution of 180~meV.

To characterize the morphology of the measured areas down to the atomic level, we observed them in a Nion UltraSTEM100 electron microscope operated at 60~keV in near-ultrahigh vacuum (beam convergence semiangle 30~mrad and medium angle annular dark field [MAADF] detector angular range 60--200 mrad). Brighter contrast in the images corresponds to greater scattering and thus greater sample thickness. The sample was baked in vacuum at 130~$^{\circ}$C for 16 h before insertion into the microscope through the ambient.

We also mapped the same areas in air using diffraction-limited confocal Raman spectroscopy (Witec Alpha300R, Witec GmbH, Germany). The focused 532 nm laser spot diameter at 1 mW power was $\sim$250 nm (-6 dB), and the sample was laterally translated using an integrated piezo stage. The scan window was 3$\times$3 $\mu$m$^2$ in size, consisting of 10 000 spectra (100$\times$100) each integrated for 500 ms. To improve the signal to noise ratio, the data was downsampled with a 6$\times$6 pixels sliding window median filter followed by background subtraction.

\begin{figure}
\includegraphics[width=0.46\textwidth]{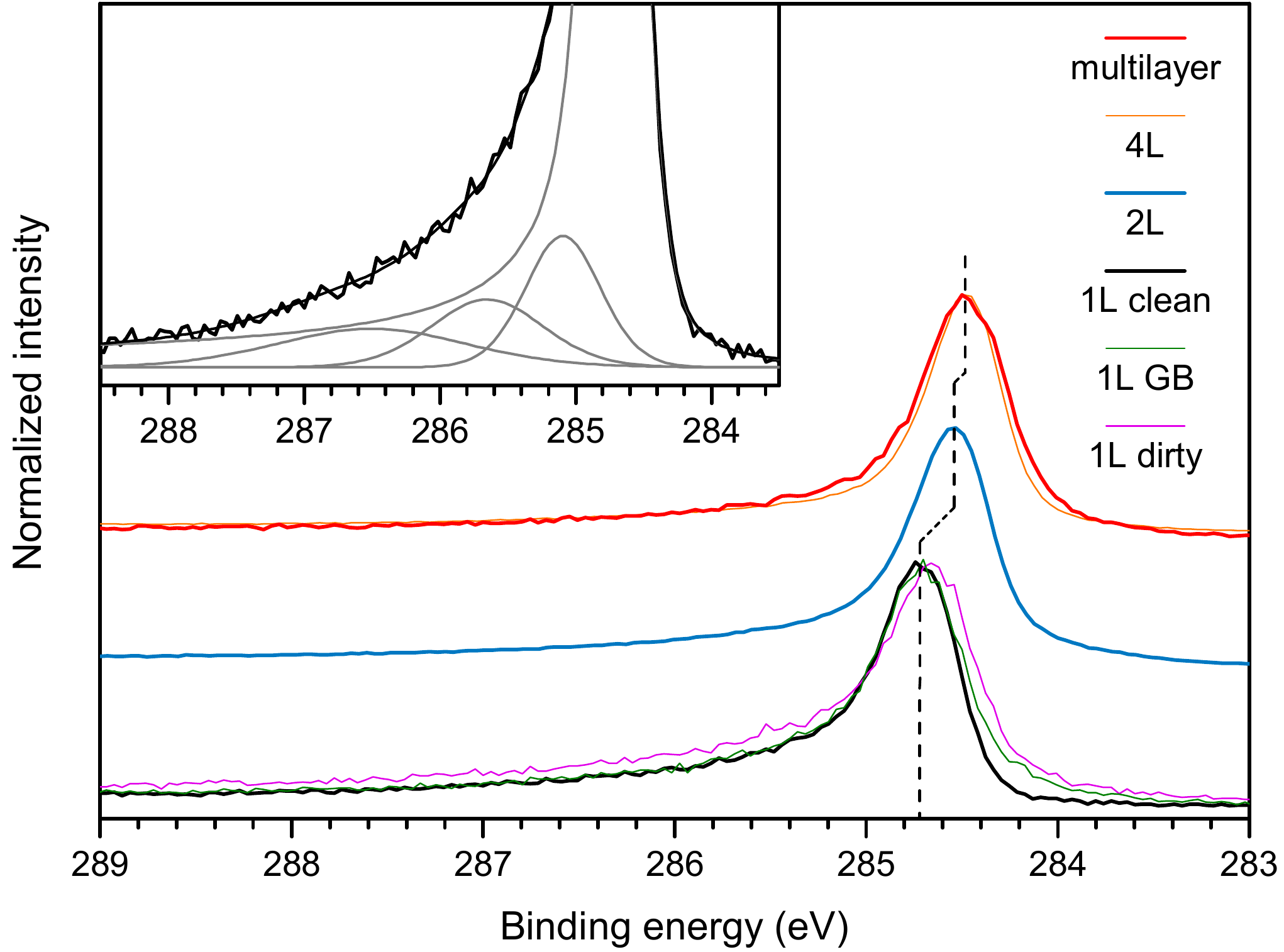}%
\caption{C~1\textit{s} XPS spectra collected from 130~nm diameter spots characterized by STEM and Raman, with the lines colored according to the color of each spot marked on the STEM images: dirty monolayer (1L) and 1L with a grain boundary (GB) in Fig.~\ref{fig:cleandirty}, clean 1L and multilayer in Fig.~\ref{fig:monomulti}), and two-layer (2L) and four-layer (4L) graphene in Fig.~\ref{fig:multilayers}. The inset gives a deconvolution of the 1L spectrum, which contains small residual contributions from non-graphitic carbon.\label{fig:spectra}}
\end{figure}

\begin{figure*}[b!]
\includegraphics[width=0.75\textwidth]{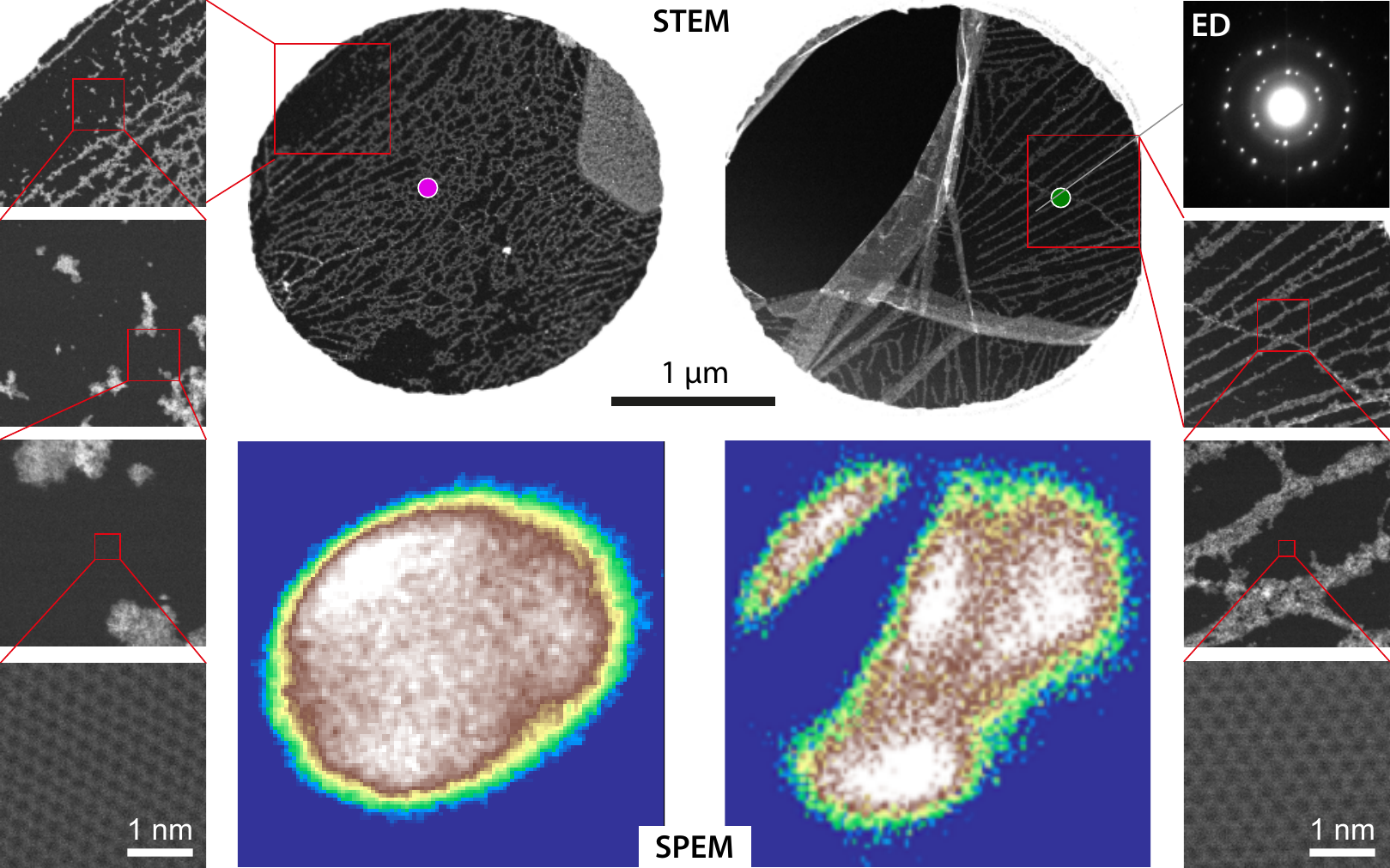}%
\caption{Morphology and spectrum maps of monolayer graphene as probed by correlated scanning transmission electron microscopy (STEM, 60 keV, MAADF detector), scanning photoelectron microscopy (SPEM, 401.03 eV photon energy, 128x128 pixels of 0.02 $\mu$m$^2$, cropped images integrated over the C~1\textit{s} response), and electron diffraction (ED, 5 kV). The colored circles approximate the size of the X-ray spot (and correspond to the spectra in Fig.~\ref{fig:spectra}) and the scale bar applies to both STEM and SPEM images, which display the same sample area.\label{fig:cleandirty}}
\end{figure*}

\begin{figure*}[t!]
\includegraphics[width=0.62\textwidth]{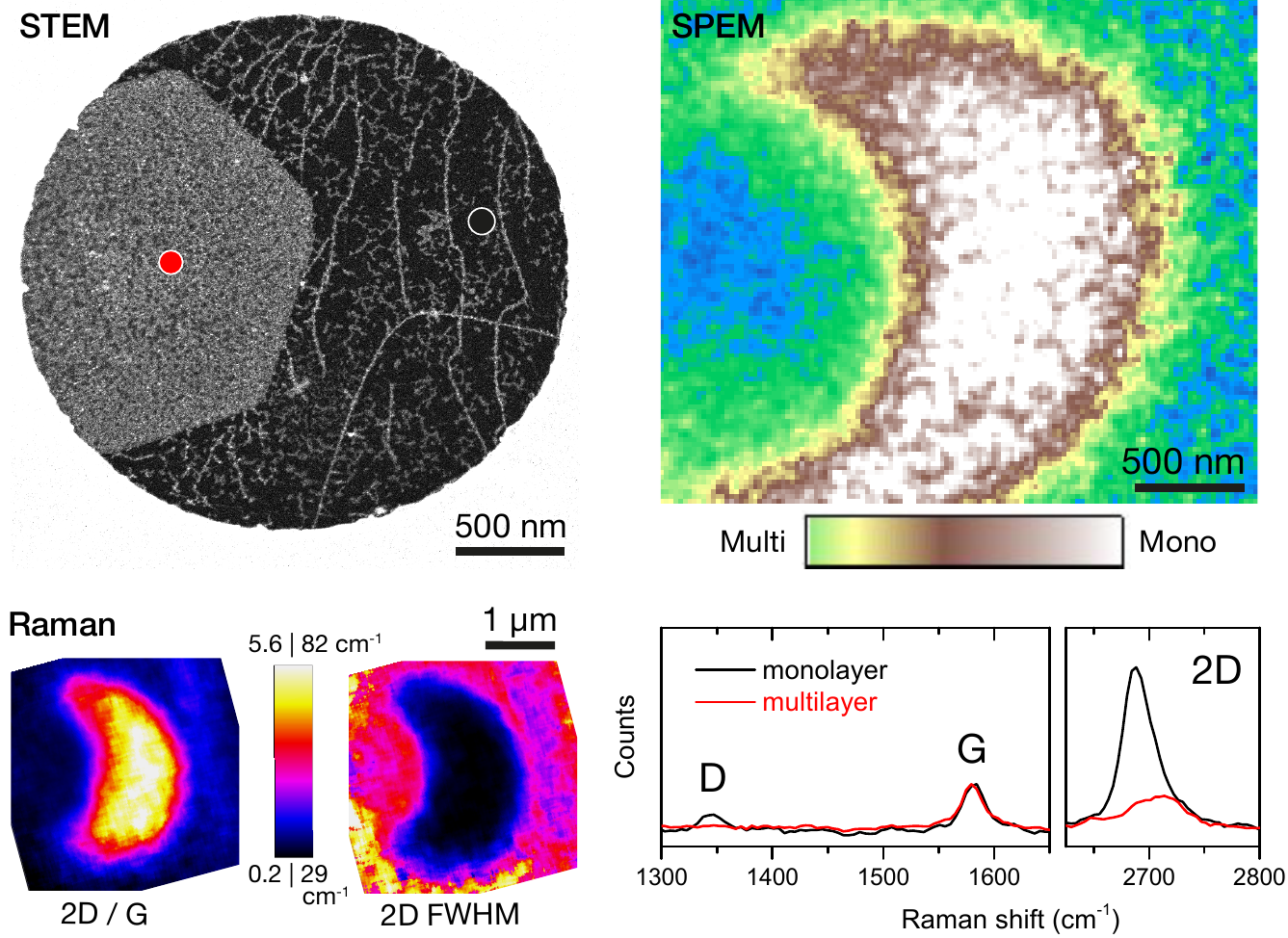}%
\caption{Monolayer graphene with an overlying multilayer grain measured using STEM, SPEM (colored according to the ratio of signal in energy windows corresponding to mono- $[284.57,284.98]$ eV and multilayer $[283.94,284.49]$ eV graphene) and Raman spectroscopy (532 nm excitation, maps of the 2D / G band intensity ratio and the 2D full width at half maximum, FWHM). The colored circles approximate the size of the X-ray spot (and correspond to spectra in Fig.~\ref{fig:spectra}). The STEM and SPEM images show the same sample area.\label{fig:monomulti}}
\end{figure*}

\begin{figure*}
\includegraphics[width=0.61\textwidth]{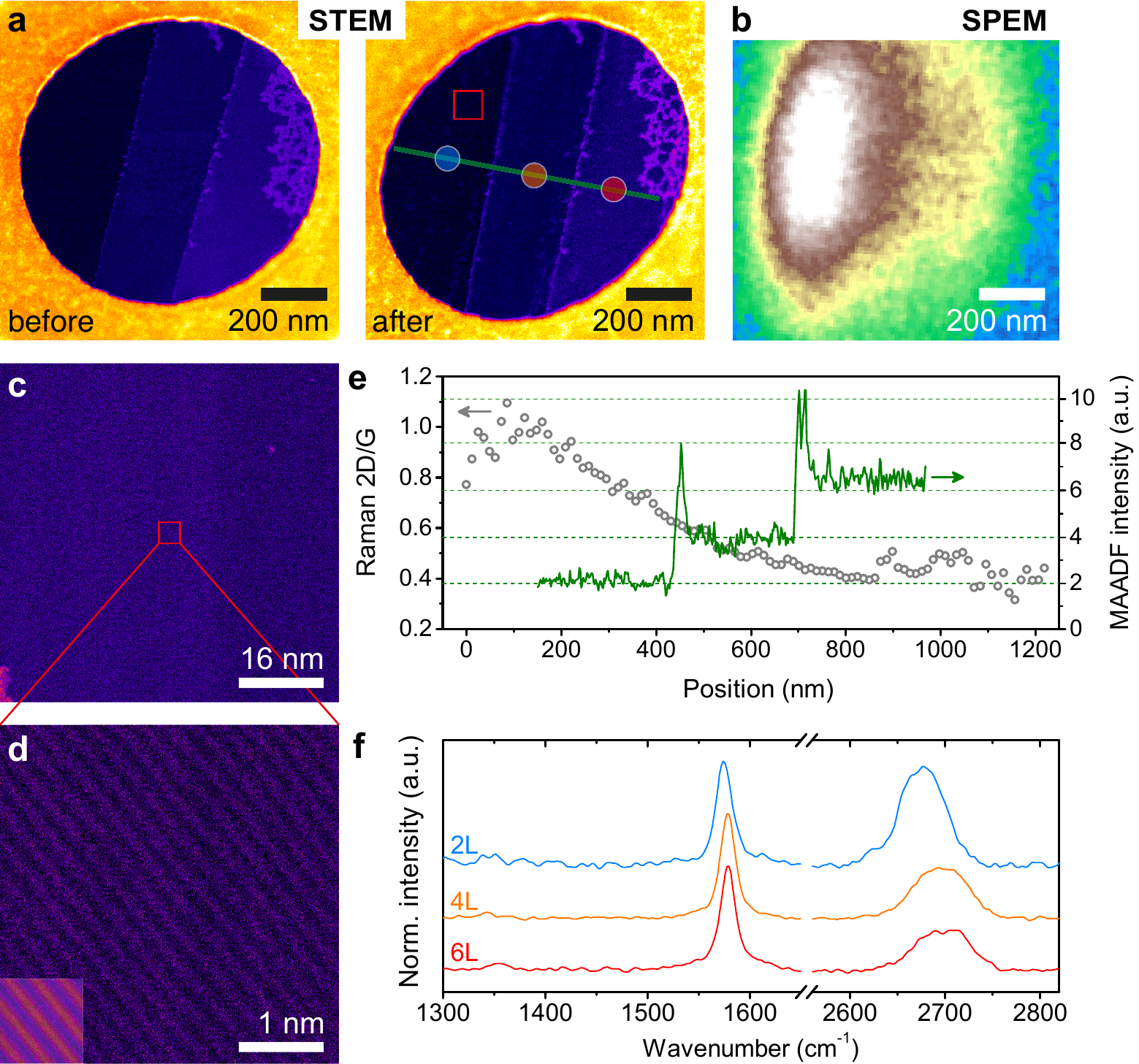}%
\caption{Characterization of few-layer graphene. a) MAADF/STEM images (with ImageJ lookup table 'Fire') acquired before and after SPEM. The oval shape of the support foil hole indicates distortion due to heat treatments and/or handling. The red open square corresponds to the closeup in panel c, the green line to the line profile in panel e, and the blue, orange and red circles to the approximate spots of the XPS (2L and 4L plotted in Fig.~\ref{fig:spectra}) and Raman spectra (panel f). b) SPEM map acquired over the same area (higher BE signal of the thinner region tends towards white). c) Large area of atomically clean lattice. d) Lattice-resolution closeup of the bilayer region. The overlaid quantitative image simulation of AB-stacked bilayer graphene tilted with respect to the electron beam by 5 and 9 degrees in x and y perfectly matches the lineal contrast. e) Raman 2D/G ratio (left axis; diffraction-limited 532 nm excitation), and the MAADF intensity line profile (right axis) showing a stepwise increase of the (vacuum background subtracted) scattering intensity, alongside a continuous decrease in the 2D/G ratio. f) Raman spectrum measured from the blue spot in panel a is consistent with bilayer graphene, with nearly no D band visible in the spectra. \label{fig:multilayers}}
\end{figure*}

Based on the X-ray photoemission contrast and the center markings of the TEM grid, we were able at the SPEM instrument to precisely identify the holes in the carbon support film over which the graphene was suspended, and later find the same areas with both STEM and Raman. In total we measured over 20 suspended areas, which mostly show similar spectral and morphological characteristics. The data allow us to disentangle three different influences on the core level signal (Fig.~\ref{fig:spectra}): remaining contamination (Fig.~\ref{fig:cleandirty}, visible as patchy brighter contrast), grain boundaries (Fig.~\ref{fig:cleandirty}, sharp bright single lines, perpendicular to lines of contamination possibly present due to wrinkling and strain), and multilayer regions (Figs.~\ref{fig:monomulti} and \ref{fig:multilayers}, uniformly brighter areas).

Figure~\ref{fig:cleandirty} shows cleaner and dirtier monolayer regions, and a boundary between two misaligned grains. This allows us to address the influence of contamination or defects (Fig.~\ref{fig:spectra}). We find neither to greatly affect the C~1\textit{s} position, only causing slight broadening without apparent increase in the asymmetry of the peak. The broadening may be explained by a convolution of the graphene spectrum with that of the residual overlying carbonaceous contaminants~\cite{Tripathi17PSSRRL}, with possible small contributions from non-sp$^2$ carbon within the graphene lattice itself.

The area where we measured the sharpest monolayer C 1\textit{s} response also included a thicker region, shown in Fig.~\ref{fig:monomulti}. A Raman spectrum map suggests this to be 4-6 layers (2D/G areal intensity ratio is $\sim$1.9 for the multilayer and $\sim$5.8 for the 1L). Interestingly, our narrow (23.1~cm$^{-1}$ full width at half maximum [FWHM]) 1L 2D peak at 2689~cm$^{-1}$ is shifted by 15-20~cm$^{-1}$ with respect to typical suspended graphene~\cite{Berciaud09NL}, and has a slight asymmetry towards the higher shift side. The core level binding energy of the monolayer is found at 284.70~eV (with a FWHM of 0.44 eV including a Gaussian width of 0.3 eV to describe thermal broadening and our energy resolution, and a Doniach-\v{S}unji\'{c} asymmetry parameter of 0.095), while for the multilayer, it is very close to that of graphite at 284.46 eV (Fig.~\ref{fig:spectra}).

These findings suggest that residual contamination or the small D band in the monolayer Raman spectrum in Fig.~\ref{fig:monomulti} (unsurprising considering the long Raman scattering activation length~\cite{Lucchese10C}), cannot be responsible for the observed large shift in the BE of the monolayer. The presence of the multilayer region and its observed BE serves as an additional reference for energy calibration, giving us full confidence in the observed monolayer value.

In addition, we characterized an exfoliated sample including a spot with stepwise varying layer number (Fig.~\ref{fig:multilayers}). We imaged the same area with STEM before and after the SPEM measurements, finding that vacuum annealing preceding spectromicroscopy was able to nearly completely clean even multilayer areas. Raman spectroscopy (2D/G areal intensity ratio of $\sim$2.5) and STEM imaging show the thinnest region to be a bilayer, although sample distortion caused by the heat treatments has induced non-planarity that prevented us from obtaining better than lattice resolution. Nonetheless, comparison of the contrast in the thinnest region to an image simulation along with the stepwise increase in MAADF intensity indicate the presence of 2, 4 and 6 layers. The SPEM map over the same region thus allows us to obtain additional XPS spectra for few-layer graphene (Fig.~\ref{fig:spectra}).

We find that the C~1\textit{s} binding energy varies nearly linearly with layer number, similar to what has been found for epitaxial graphene~\cite{Hibino09PRB}. Since our samples are suspended, however, we can be confident that these shifts are due to intrinsic differences in the screening of the core hole, and not affected by a substrate. Via concurrently measured valence band photoemission, we found no shift of the Dirac point within our energy resolution (sufficient to distinguish shifts on the order of 100 meV, as observed for the C 1\textit{s}). Suspended multilayers are also possible to unambiguously model, allowing us to turn to density functional theory simulations.

To understand the physical origin of this shift, we calculated the C~1\textit{s} BEs of monolayer and slabs of AB-stacked bi-, tri- and four-layer graphene using total energy differences implemented via core projectors~\cite{Ljungberg11JESRP,Susi15PRB} in the \textsc{Gpaw} package~\cite{Enkovaara2010}. This level of theory is able to approximate the static screening of the core hole by the conduction electrons (potentially also from neighboring layers). In each case, we used large 5$\times$9 supercells of the orthorhombic 4-atom unit cell to minimize spurious interaction between the periodic images of the core hole~\cite{Susi15PRB}. We tested several exchange-correlation functionals: in addition to the local density approximation (LDA)~\cite{Perdew92PRB} and the Perdew-Burke-Ernzerhof (PBE)~\cite{Perdew96PRL} generalized gradient approximation, two functionals based on nonlocal van der Waals correlations (vdW-DF family, with the spin generalization based on~\cite{Thonhauser15PRL}) from the libvdwxc library~\cite{Larsen17MSMSE}, vdW-DF2~\cite{Lee10PRB} and C09-vdW~\cite{Cooper10PRB}.

To obtain the geometries, we first relaxed~\cite{Larsen17JPCM} orthorhombic four-atom unit cells of single- and few-layer graphene with a plane-wave basis (600~eV cutoff energy, 12 \AA\ of perpendicular vacuum, 9$\times$15$\times$1 Monkhorst-Pack \textbf{k}-points, maximum force $<$0.0002~eV/\AA). We then calculated the total energy difference $E_B$ between each ground and first core-excited states (finite-difference grid spacing 0.19~\AA, 3$\times$3$\times$1 \textbf{k}-points) in a square 5$\times$9 supercell with 180 C atoms per layer (up to 720 for the four-layer cell). These cell sizes and \textbf{k}-point meshes yielded BEs converged within a few tens of meV, and within a few meV for their differences.

The calculated BEs for each layer number are listed in Table~\ref{tab:BEs}. We find that, apart from expected differences in absolute values~\cite{Susi15PRB}, all functionals predict the correct trend, with a downshift from monolayer to four layers of around 100 meV. The magnitude of this shift is less than half of our experimental value, regardless of whether van der Waals interactions were explicitly included, and regardless of the relaxed layer spacing. This indicates that the screening effect of adjacent layers is well described by charge rearrangements within each layer.  However, the static screening included in our model only accounts for part of the full screening effect, suggesting that higher levels of theory will be required for further accuracy.
 
\begin{table}[t!]
\caption{Calculated core level binding energies $E_B$ for monolayer (1L), bilayer (2L), trilayer (3L) and four-layer (4L) graphene and their average downshift $\Delta E_B = E_B^{1L} - (E_B^{nL,A} + E_B^{nL,B})/2 $ (where A and B denote values for the two graphite sublattices, which vary by $<$20 meV) for the various density functionals (vdW label omitted from DF2 and C09 for brevity), along with the relaxed 2L layer spacing $d$.
\label{tab:BEs}}
\begin{ruledtabular}
\begin{tabular}{ccccccc}
Functional & LDA & PBE & DF2 & C09 & Exp. \\
\colrule
d (\AA) & 3.371 & 3.758 & 3.568 & 3.314 & 3.27\footnote{Ref.~\citenum{Brown12NL}, $\pm$0.18 \AA.},  3.46\footnote{Ref.~\citenum{Nicotra13AN}, $\pm$0.13 \AA.} \\
$E_B^{1L}$ (eV) & 280.636 & 283.662 & 286.577 & 283.162 & 284.70 \\
$\Delta E_B^{2L}$ (meV) & -38 & -51 & -48 & -28 & -160 \\
$\Delta E_B^{3L}$ (meV) & -83 & -64 & -80 & -65 & -- \\
$\Delta E_B^{4L}$ (meV) & -118 & -92 & -103 & -124 & -230 \\
\end{tabular}
\end{ruledtabular}
\end{table}

To conclude, by correlating electron microscopy with Raman spectroscopy and synchrotron-based scanning photoelectron microscopy, we have for the first time measured the core level photoemission response of comprehensively characterized suspended mono- and few-layer graphene areas. We find that contamination or grain boundaries slightly broaden the signal, whereas the core level binding energy of the monolayer is upshifted by 280~meV from the value in graphite, with intermediate values found for two- and four-layer graphene.

While the observed shifts are in good agreement with those measured for few-layer epitaxial graphene~\cite{Hibino09PRB}, any small Dirac point variations in suspended monolayers due to charge transfer doping~\cite{Knox11PRB} cannot here explain the magnitude of the shift. Further improvements in modeling are required to precisely reproduce the photoelectron signal, although our close agreement between all our functionals indicates that van der Waals interactions between adjacent layers do not play a significant role.

\begin{acknowledgments}
T.S. acknowledges the Austrian Science Fund (FWF) for funding via project P 28322-N36, the European Research Council (ERC) Grant No. 756277-ATMEN, and the Vienna Scientific Cluster for computational resources. M.S. is a FNRS Postdoctoral researcher. A.H.L. acknowledges funding from the European Union's Horizon 2020 research and innovation program under grant agreement no.~676580 with The Novel Materials Discovery (NOMAD) Laboratory, a European Center of Excellence; the European Research Council (ERC-2010-AdG-267374); Spanish grant (FIS2013-46159-C3-1-P); and Grupos Consolidados (IT578-13). K.M., A.M., C.M. and J.C.M. acknowledge funding by the ERC Grant No. 336453-PICOMAT, and J.C.M by the FWF project P25721-N20. J.K. was supported by the FWF project I 3181-N36 and the Wiener Wissenschafts\mbox{-,} Forschungs- und Technologiefonds (WWTF) via project MA14-009.
\end{acknowledgments}

%

\end{document}